\documentstyle[12pt,aaspp]{article}

\def\hangpar{\par\noindent\hangindent=7mm\hangafter=1}
\def\numophbinaries{30}
\def\numtaubinaries{55}
\def\numophsingles{47}
\def\pms{pre--main-sequence}
\def\msun{\hbox{${M}_{\sun}$}}

\def\multiplemark{T}

\slugcomment{To appear in the {\it Astrophysical Journal}, Feb.\ 10, 1996.}

\begin{document}

\title{The Connection between Submillimeter Continuum\\ Flux and
 Binary Separation in Young Binaries:\\ Evidence of Interaction
 between Stars and Disks}

\author{Eric L. N. Jensen\altaffilmark{1} and Robert D.
  Mathieu\altaffilmark{2}}
\affil{Department of Astronomy, University
  of Wisconsin-Madison, Madison, WI 53706}

\and

\author{Gary A.  Fuller\altaffilmark{3}}
\affil{NRAO, Edgemont Road, Charlottesville, VA 22903}

\altaffiltext{1}{E-mail: jensen@astro.wisc.edu}
\altaffiltext{2}{E-mail: mathieu@astro.wisc.edu}
\altaffiltext{3}{Previously at NRAO, 949 N. Cherry Ave., Campus
  Building 65, Tucson, AZ 85721. E-mail: gfuller@nrao.edu.  The
  National Radio Astronomy Observatory is operated by Associated
  Universities, Inc., under a cooperative agreement with the National
  Science Foundation.}

\begin{abstract}
  We present 800 \micron\ continuum photometry of \pms\ binary stars
  with projected separations $a_p < 150$ AU in the Scorpius-Ophiuchus
  star-forming region.  Combining our observations with published 1300
  \micron\ continuum photometry from Andr\'e \& Montmerle (1994), we
  find that binaries in Sco-Oph with $1 < a_p < $ 50--100 AU have
  lower submillimeter continuum fluxes than wider binaries or single
  stars, as previously found for Taurus-Auriga binaries.  The wide
  binaries and single stars have indistinguishable submillimeter flux
  distributions. When the Sco-Oph and Tau-Aur samples are combined,
  this dependence of submillimeter flux on binary separation is
  detected with a confidence level of greater than 99\%.  Thus,
  binary companions with separations less than 50--100 AU significantly
  influence the nature of associated disks.

  We have explored the hypothesis that the reduction in submillimeter
  flux is the result of gaps cleared in 100-AU disks by companions.
  Gap clearing produces the qualitative dependence of submillimeter
  flux on binary separation, and a simple model suggests that large
  gaps in disks with surface densities typical of wide-binary or
  single-star disks can reduce submillimeter fluxes to levels
  consistent with the observed limits. This model shows that the
  present submillimeter flux upper limits do not necessarily imply a
  large reduction in disk surface densities.

  Two-thirds of the \pms\ binaries were detected by IRAS at 60
  \micron, showing that most binaries have at least one circumstellar
  disk.  We have used these fluxes to place lower limits of $10^{-5}$
  \msun\ on circumstellar disk masses. Similarly, the 60 \micron\
  fluxes indicate that the circumstellar disk surface densities are no
  more than two orders of magnitude smaller than those of typical
  disks around single stars.

  Our upper limits on submillimeter fluxes place upper limits of 0.005
  \msun\ on circumbinary disk masses. Thus massive circumbinary disks
  (such as that found around GG Tau) are rare among binaries with
  projected separations between a few AU and 100 AU\null.
  Circumbinary disks are found around some binaries with separations
  less than a few AU\null.

\end{abstract}


\section{Introduction}\label{sec:intro}

It has been known for some time that a majority of main-sequence stars
are in binary systems (e.g.\ Abt 1983, Duquennoy \& Mayor 1991).
Recently, systematic surveys of \pms\ (PMS) stars have shown that
their binary frequency is at least as high as, and perhaps
significantly higher than, that of main-sequence stars (Mathieu,
Walter, \& Myers 1989, Simon et al.\ 1992, 1995, Ghez, Neugebauer, \&
Matthews 1993, Leinert et al.\ 1993, Reipurth \& Zinnecker 1993).
Thus, the most likely outcome of the star formation process is a
binary system and understanding binary formation is vital to
understanding star formation.

Infrared and millimeter observations have shown that many young stars
have excess long-wavelength emission, indicative of circumstellar
material.  Although this material is unresolved in most observations,
there is mounting evidence that in most cases it lies in circumstellar
or circumbinary disks (see e.g.\ Beckwith \& Sargent 1993, Basri \&
Bertout 1993, and Sargent 1995 for reviews).  Indeed, much of a star's
mass may be built up by accretion of material through these disks.
Thus, disks provide a fossil record of the star formation process, and
study of disk properties may reveal clues about star formation.

Observations suggest that the structure of a disk is affected by the
presence of an embedded binary.  Beckwith et al.\ (1990, hereafter
BSCG) measured 1300 \micron\ continuum emission from 86 PMS stars in
Taurus-Auriga (Tau-Aur) to search for disks.  They found that very few of the
binaries in their sample with projected separations less than 100 AU
had detectable 1300 \micron\ emission while strong emission was common
in wider binaries (see also Beckwith \& Sargent 1993).  Jensen,
Mathieu, \& Fuller (1994, hereafter Paper I) used a larger sample of
binaries in Tau-Aur and more sensitive 800 \micron\ observations
of the systems with the smallest projected separations to further
explore this result.  We found that the fluxes from binaries with
projected separations between 1 and 50--100 AU were significantly
lower than those from wider binaries or single stars.  On the other
hand, the flux distributions of the two latter groups are
statistically indistinguishable.  Osterloh \& Beckwith (1995) made
additional 1300 \micron\ observations of PMS stars
in Tau-Aur and did detect some close binaries.  Nonetheless,
they found the difference in flux distributions between close binaries
and wide binaries or single stars to remain at a statistically
significant level.

There are a variety of theoretical predictions about the influence of
binaries on their associated disks.  A binary companion embedded in a
disk is expected to rapidly clear a gap, isolating distinct
circumstellar and circumbinary disks.  Such a gap may inhibit the
transfer of material from circumbinary to circumstellar disks (e.g.\
Artymowicz et al.\ 1991, but see Artymowicz \& Lubow 1994).  If so,
continued accretion from the circumstellar disks onto the stars may
exhaust the reservoir of circumstellar disk material more quickly than
in single stars (Clarke 1992).  Indeed binary companions may
accelerate the depletion.  Ostriker, Shu, \& Adams (1992) found that
for separations less than 100 AU a companion can excite density waves
in a circumstellar disk, causing an enhanced accretion rate onto the
central star.  At the same time, the inhibition of accretion could
increase the lifetimes of circumbinary disks (Clarke 1992).  Pringle
(1991) found that the transfer of angular momentum from the binary
orbit to the circumbinary disk pushes disk material to larger radii,
resulting in a reduction in surface density and increase in size of
the circumbinary disk.  We note that most of these predictions remain
largely untested by current observations.

We have undertaken a submillimeter survey of young binaries in the
Scorpius-Ophiuchus (Sco-Oph) star-forming region.  In this paper, we
present 800 \micron\ observations of most known Sco-Oph PMS binaries
with projected separations $a_p$ less than 150 AU\null. We combine
these data with other Sco-Oph submillimeter data from Andr\'e \&
Montmerle (1994) and find a dependence of submillimeter flux on binary
separation in a sample that is independent of that used by BSCG, Paper
I, or Osterloh \& Beckwith (1995).  We compare the submillimeter data
from Sco-Oph and Tau-Aur with submillimeter fluxes predicted from
models of disks with gaps and show that gap clearing alone may be
sufficient to explain the low submillimeter fluxes from the close
binaries. Most PMS binaries at all separations were detected at 60
\micron\ by IRAS, indicating the presence of circumstellar disks.  We
derive lower limits on circumstellar disk masses from the 60 \micron\
emission. Finally, we use submillimeter fluxes to place upper limits
of 0.005 \msun\ on circumbinary disk masses for typical binaries with
$a_p$ between a few AU and 100 AU.

\section{Observations}

The target binaries (see \S\ref{section:sample}) were observed using
the James Clerk Maxwell Telescope (JCMT)\footnote{The James Clerk
  Maxwell Telescope is operated by the Royal Observatories on behalf
  of the Particle Physics and Astronomy Research Council of the United
  Kingdom, the Netherlands Organization for Scientific Research, and
  the National Research Council of Canada.} on 1992 February 24--27,
1992 March 2--4, and 1993 February 26 -- March 1.  The JCMT facility
$^3$He cooled bolometer, UKT14, was used for the observations.  All
observations were made using a 65-mm aperture which resulted in a
$\sim19\arcsec$ beam width (FWHM) at all wavelengths. All the
observations were made using standard chopping and beam switching with
synchronous detection of the signal. The secondary chopper was
switched 60\arcsec\ in azimuth at a rate of 7.8 Hz.

During the 1992 observing sessions the atmosphere was usually stable
and transparent.  Measurements from a tipping radiometer indicated
zenith optical depths of typically 0.03 to 0.04 at 225 GHz.  The
weather during the 1993 run was very stable and the atmosphere
transparent. The zenith optical depth at 225 GHz was approximately
constant at $\sim0.08$ during the first night and between 0.04 and
0.05 for the remaining nights.  During both runs standard sources of
known flux were observed every 15 to 30 minutes in order to monitor
the sky opacity.  The pointing of the telescope was also checked and
corrected regularly during the observations by making five-point
measurements on bright sources.  Typically the actual pointing was
found to be within $\sim3$\arcsec\ of the nominal pointing. Very
occasional shifts as large as 6\arcsec\ were seen.

Our survey observations were made at 800 \micron.  Because thermal
dust emission decreases with wavelength from 800 to 1300 \micron\ more
quickly than the atmospheric transmission improves (in good weather),
at the JCMT 800 \micron\ observations are more sensitive for detecting
weak dust emission. In addition, non-thermal emission (e.g.\ free-free
or gyrosynchrotron) increases in strength with increasing wavelength,
and thus 800 \micron\ observations are less likely to be contaminated
by this non-disk emission.

The program sources were initially observed for $\sim 1000$ s at 800
\micron.  This resulted in a typical 1$\sigma$ noise level of 15--20
mJy. Most sources were observed twice, particularly those for which
the noise level of the first observation was higher than the typical
level of sources which were marginally detected. Observations at 1100
\micron\ and 450 \micron, and occasionally 350 \micron, were made of
sources which were strong detections at 800 \micron.

Observations of the standard sources at different airmasses were used
to determine the telescope sensitivity at 1100 \micron, 800 \micron,
450 \micron, and 350 \micron.  Given the telescope sensitivity, the
standard source observations provided an estimate of the sky opacity
as a function of time throughout each night. For both observing
sessions, the variation of the sky opacity with time derived from
these observations very closely tracked the opacity derived from the
225 GHz radiometer.  The program sources were calibrated using the
mean sky opacity derived from standard sources observed before and
after the target source.  The scatter in the zenith opacity about a
linear fit versus time for each night has been used as an estimate of
the uncertainty in the flux calibration.

Each observation consisted of a series of 10-second samples.  To
remove noise spikes from the data, the mean of the samples in each
integration was calculated and those samples which deviated from the
mean by more than three standard deviations were removed and the mean
was recalculated.  The measured fluxes or $3\sigma$ upper limits are
given in Table \ref{table:jcmt_data}.  The quoted uncertainties are a
combination of the flux calibration uncertainty discussed above and
the photon noise derived from the standard deviation of the individual
10-second samples within each observation.

\section{Sample selection}\label{section:sample}

We chose binaries for our sample from among members of the Upper
Scorpius subgroup of the Sco OB2 association and from the $\rho$ Oph
cloud complex.  These stars lie within the approximate boundaries of right
ascension 15$^{\rm h}$10$^{\rm m}$--16$^{\rm h}$40$^{\rm m}$ and
declination $-5\deg$--$-35\deg$ (Blauuw 1978).  We will refer to this
area as ``Sco-Oph''.  We adopt distances of 125 pc for the $\rho$ Oph
clouds and 160 pc for Upper Sco (de Geus, de Zeeuw, \& Lub 1989).

Surveys which have searched for binaries in Sco-Oph include those of
Mathieu et al.\ (1989), Ghez et al.\ (1993), Reipurth \& Zinnecker
(1993), and Simon et al.\ (1992, 1995).  These surveys vary greatly in
their target lists and the range of binary separations to which they
are sensitive.  The characteristics of the surveys are summarized in
Table \ref{table:binary_surveys}.  In addition, several PMS binaries
have been found by observations that were not part of surveys.  AK Sco
was discovered to be a spectroscopic binary by Andersen et al.\
(1989).  WL 2 and WL 20 were both found to be binaries by infrared
imaging (Rieke, Ashok, \& Boyle 1989, Barsony et al.\ 1989). VV Sco
was noted as a visual binary by Gregorio-Hetem et al.\ (1992).  DoAr
51 was found to be a binary by Koresko (1995) using speckle
interferometry.

Currently only spectroscopic observations can detect binaries with
separations less than $\sim$ 1 AU at the distance of Sco-Oph.  All but
one (AK Sco) of the known spectroscopic binaries in Sco-Oph were found
in a radial-velocity survey of x-ray selected stars in Sco-Oph
(Mathieu et al.\ 1989). Independent of multiplicity, most of the x-ray
selected stars surveyed by Mathieu et al.\ (1989) do not have infrared
excesses or strong H$\alpha$ emission (Walter et al.\ 1994), the
classical signatures of disks.  Thus the binaries discovered in that
survey may not be representative of PMS spectroscopic binaries in
general, particularly in terms of the disk properties with which we
are concerned here.

In contrast to the spectroscopic observations of Mathieu et al.\
(1989), the other surveys for binaries in Sco-Oph have drawn their
target lists primarily from lists of H$\alpha$- and infrared-selected
PMS stars.  Ghez et al.\ (1993) and Simon et al.\ (1995) also include
some x-ray selected PMS stars in their target lists.  The targets of
these surveys were not initially discovered to be young stars using a
uniform set of selection criteria, and thus the combined sample of
Sco-Oph binaries from these surveys is not a well-defined,
uniformly-selected PMS binary sample.  However, the diversity of
targets of these surveys reflects the variety of techniques that have
been used to search for young stars.  Thus, while the sample of
binaries discovered by these surveys may not be uniformly defined, it
should be reasonably representative of the known population of PMS
stars in Sco-Oph in general.

Because the spectroscopic binaries are drawn from a distinctly
different sample of PMS stars than are the other binaries, in this
paper we limit our analysis to binaries with projected separations
greater than 1 AU\null.  This sample should be neither more nor less
biased toward the presence of disks than the known Sco-Oph PMS
population as a whole.  For these binaries, we present a quantitative
analysis of the distribution of submillimeter fluxes with projected
separation and discuss the implications of our data for binary-disk
interactions.  We will discuss the submillimeter fluxes and spectral
energy distributions of binaries with projected separations less than
1 AU in a later paper.

Simon et al.\ (1992, 1995) argue that for separations less than
10\arcsec, contamination of the binary sample from chance
superposition of stars should be negligible.  Thus, we adopt this as
the separation upper bound for our sample.  We have taken projected
separations or projected semimajor axes for Sco-Oph binaries from the
binary surveys discussed above.

The other primary requirement for inclusion in our sample is
availability of an 800 \micron\ or 1300 \micron\ flux measurement.
(For convenience, we will refer to either of these as ``submillimeter
fluxes.'') Submillimeter fluxes or flux limits are from this work and
Andr\'e \& Montmerle (1994, hereafter AM), which together cover about
90\% of the currently known PMS binaries in Sco-Oph.

Our goal is to explore the influence of multiplicity on the properties
of circumstellar and circumbinary disks.  In order to avoid confusion
introduced by other factors that may also influence submillimeter
emission, we have further constrained our sample.
Our sample is limited to PMS stars with spectral types of F
and later so as to reduce the range of
stellar mass, luminosity, and temperature.
We have also excluded Class I sources as defined in the infrared
classification scheme of Lada (1987).  Terebey, Chandler, \& Andr\'e
(1993) suggest that a significant fraction of the 1300 \micron\
emission from Class I sources arises in an envelope rather than a
disk; this is consistent with the fact that Class I sources have
significantly higher 1300 \micron\ fluxes on average than Class II or Class
III sources and that Class I sources tend to be extended whereas Class
II sources are unresolved (AM).  Class I sources are also heavily
embedded and presumably younger than Class II or III sources (AM).
In fact, the spectral type and infrared class criteria remove only a few
systems from our sample.

After applying these selection criteria, we obtain a sample of
\numophbinaries\ multiple systems in Sco-Oph.  This is comparable to
the 42 systems in Tau-Aur studied in Paper I\null. With the addition
of data from Osterloh \& Beckwith (1995), the present Tau-Aur sample
includes \numtaubinaries\ multiple systems.  We also analyze this
expanded sample below.

In addition to the binaries, we adopt a sample of single stars for
comparison.  Single stars provide a ``control'' sample, revealing
submillimeter fluxes produced in the absence of any stellar-mass
companion.  We take our single-star sample from the targets observed
by AM which are not known to have stellar companions within 10\arcsec.
We apply the same spectral type and IR class criteria described above.
 From the $\rho$ Oph region, we adopt only the sources that AM have
marked with a ``\#'' in their Table 1, since these are all confirmed
PMS stars which have been observed at a uniform sensitivity.  In
addition, we adopt the other single PMS stars which AM observed that
lie outside $\rho$ Oph but within Sco-Oph.  This yields a total of
\numophsingles\ stars.

We note that this definition of our single star sample is less
restrictive than in Paper I for the Tau-Aur sample, where we adopted as
single only those stars which had been surveyed using speckle
interferometry or lunar occultation and found not to have companions.
However, the surveys for binaries in Sco-Oph have been less
comprehensive and less uniform than in Tau-Aur; applying such a
criterion to Sco-Oph would yield a prohibitively small sample of 15
stars.  Thus, here we adopt as single those stars not known to have
companions, whether or not they have been surveyed with
high-resolution techniques.  To estimate the contamination from
unresolved companions in our single-star sample, we note that of
\numophsingles\ stars in the ``single'' sample, 15 have been surveyed
using high-resolution techniques, leaving 32 of unknown multiplicity.
Adopting a multiplicity frequency of 50\% (Simon et al.\ 1995 find
48\% for Tau-Aur and Oph), we then expect $\sim 16$ binaries out of
\numophsingles\ stars in the sample.  Wide companions ($a_p \gtrsim
1\arcsec$) would be more likely to have been previously discovered, so
the contamination is likely to be predominantly close binaries.  Thus
the expected effect of contamination of the single sample by
undetected binaries would be to weaken the statistical significance of
any differences between the flux distributions of the close binaries
and single stars.

\section{The dependence of submillimeter flux on binary separation}
\subsection{Combining 1300 \micron\ and 800 \micron\ flux measurements}

The bulk of the submillimeter and millimeter continuum flux
measurements of PMS stars in the literature are contained in the
surveys of BSCG, Osterloh \& Beckwith (1995) (both including Tau-Aur),
and AM (Sco-Oph).  All of these surveys were made at a wavelength of
1300 \micron\ while our observations were made at 800 \micron.
Combining these measurements raises the
question of how to compare fluxes measured at 1300 \micron\ with those
measured at 800 \micron.  Because of the existence of a large number
of 1300 \micron\ flux measurements of PMS stars in the literature, we
have chosen to scale our 800 \micron\ fluxes or flux limits to 1300
\micron\ for comparison with existing data.  Submillimeter and
millimeter fluxes from dust around PMS stars are dominated by emission
on the Rayleigh-Jeans tail of the Planck function, and thus if the
emission is optically thick the flux density $F_\nu$ should vary with
frequency as $\nu^2$.  If the emission is optically thin and the
submillimeter dust opacity varies as $\kappa_{\nu} \propto \nu^\beta$,
then the flux should vary as $\nu^{2+\beta}$.  Since dust opacity is
thought to increase with increasing frequency (i.e.\ $\beta$ is
positive; Beckwith \& Sargent 1991, Pollack et al.\ 1994, Agladze et
al.\ 1994), then the variation of submillimeter and millimeter thermal
continuum emission with frequency should be {\em at least\/} as steep
as $F_\nu \propto \nu^2$.  In fact, observations of PMS stars at
multiple millimeter wavelengths bear this out.  Beckwith \& Sargent
(1991) observed 25 PMS stars and found that 19 out of 25 have
millimeter spectral slopes greater than or equal to $\nu^2$ within
their 1$\sigma$ uncertainties.

Thus, for systems with upper limits on their 800 \micron\ fluxes we
calculate 1300 \micron\ upper limits by multiplying the 800 \micron\
flux limits by $(\nu_{1300}/ \nu_{800})^2 = 0.38$.  Since the true
spectral slope may be steeper than $\nu^2$, this scaling provides a
conservative but secure upper limit on the 1300 \micron\ flux.  For
systems with upper limits at both 800 and 1300 \micron, the measured
1300 \micron\ limit is compared to the 800 \micron\ limit scaled to
1300 \micron, and the lower of the two values is used.

One system in our Sco-Oph sample was detected at 800 \micron\ but not
1300 \micron.  Haro 1-4 was not detected by AM with a 1300-\micron\
flux limit of $F_\nu < 50 $ mJy.  This limit is consistent with our
800 \micron\ detection of $F_\nu = 58 $ mJy.  To estimate the 1300
\micron\ flux of Haro 1-4, we extrapolated the 800-\micron\ flux to
1300 \micron\ assuming a spectral slope of $\nu^2$, the most common
slope found by Beckwith \& Sargent (1991).  Nonetheless, the unknown
spectral slope leads to an uncertain 1300 \micron\ flux.  However,
tests using different slopes to determine the flux of Haro 1-4, or
leaving it out of the sample altogether, show that the effect of this
uncertainty on our results is negligible.

A few binary systems do not show the variation of submillimeter flux
with wavelength expected for thermal emission from dust.  We treat
these on a case-by-case basis.  V773 Tau was detected at 1300 \micron\
by BSCG with a flux of $42 \pm 6$ mJy and by Osterloh \& Beckwith
(1995) with a flux of $24 \pm 4$ mJy.  However, we did not detect it
at 800 \micron\ with a $3\sigma$ upper limit of 29 mJy (Paper I).  The
apparent variability at 1300 \micron, the 800--1300 \micron\ spectral
slope much flatter than $\nu^2$, and the strong non-thermal centimeter
wavelength emission (Phillips, Lonsdale, \& Feigelson 1991) suggest
that at least some of the 1300 \micron\ flux from V773 Tau is produced
by a non-thermal process and is not emission from circumstellar
dust.  As in Paper I, we proceed on the assumption that the 1300
\micron\ emission is non-thermal, and we use the 800 \micron\ limit
from Paper I to derive an upper limit on the 1300 \micron\ disk emission.

Two binaries in Sco-Oph (SR 9 and DoAr 24E) show a similar
disagreement of 800 \micron\ and 1300 \micron\ fluxes.  We did not
detect SR 9 at 800 \micron\ with a $3\sigma$ upper limit of 28 mJy.
Scaling this to 1300 \micron\ by $\nu^2$ yields an upper limit of 11
mJy, whereas AM list it as a weak detection ($\sim 4\sigma$) at 1300
\micron\ with a flux of 15 mJy.  However, in this case there is no
independent evidence of non-thermal emission; SR 9 was undetected at 6
cm (Andr\'e et al.\ 1990).  SR 9 may be similar to the few systems
found by Beckwith \& Sargent (1991) to have submillimeter spectral
slopes slightly flatter than $\nu^2$ (i.e.\ $-1 < \beta < 0$).
Because of the lack of other evidence for non-thermal contributions to
the millimeter flux and the relatively modest deviation from expected
dust-emission spectral slope, for SR 9 we adopt the 1300 \micron\ flux
measured by AM for our analysis.  DoAr 24E, however, shows a larger
disagreement between 800 and 1300 \micron\ fluxes.  We measured a flux
of $37\pm 8$ mJy at 800 \micron, while AM measured 65 mJy at 1300
\micron.  These measurements imply $\beta \approx -3.2$, well outside
the range of $\beta$ values found by Beckwith \& Sargent (1991).
Reipurth \& Zinnecker (1993) report a 1300 \micron\ flux from DoAr 24E
of $\sim 20$ mJy.  Although no radio continuum measurements of DoAr
24E are known to us, its unusual submillimeter spectral slope and
apparent variability (as in the case of V773 Tau) suggest a
non-thermal contribution to its 1300 \micron\ flux.  Thus, we adopt
our 800 \micron\ flux scaled to 1300 \micron\ by $\nu^2$ as the best
estimate of the disk contribution to submillimeter emission for the
analysis below.  Finally, we emphasize that, given the size of our
sample, the particular choice of flux for these three systems (or even
their inclusion or omission) does not affect the conclusions of our
analysis.

\subsection{Assignment of flux in unresolved triple and quadruple systems}

Five systems in our Sco-Oph sample each comprise three or four stars
that lie within the JCMT beam: V853 Oph, SR 24, 155913$-$2233,
162814$-$2427 (ROX 42C), and 162819$-$2423 (ROX 43A/B).  In the Tau-Aur binary
sample, there are six systems which are triples with all components
within 10\arcsec\ of each other: HP Tau/G3, V807 Tau, HV Tau, RW Aur,
UZ Tau, and UX Tau.  In the absence of spatially-resolved flux
measurements of the individual components of such higher-order
multiple systems, it is unknown whether any detected flux should be
associated with the binary, the tertiary, or both.
Thus the positions of such
systems in a plot of flux vs.\ projected separation are ambiguous.

For systems that are not detected, we know that neither the close pair
nor the wide companion has substantial submillimeter flux; thus the
system is reasonably represented by two points, one at each projected
separation.\footnote{However, a counter argument to this is that a
  hierarchical quadruple system is quite different from just a wide
  binary, since this ``wide pair'' would not be expected to have a
  large flux if both close binaries cleared their disks.  The only
  quadruple in our sample is 162819$-$2423 (ROX 43A/B), which is
  represented by three data points, one at each projected separation
  in the system.  However, the closest pair in the system (the
  spectroscopic binary 162819$-$2423S) has a separation smaller than
  the 1 AU lower bound of our sample.  Thus only two points for this
  system appear on the plots and in the statistical analysis. } For
detected systems, the flux could arise from the close pair, the wide
companion, or both.  For these systems we have also plotted two
points, one at each projected separation.  For the pair with the
larger projected separation, at least one member (whether that member
is single or multiple) of the wide pair has detectable submillimeter
flux.  Thus, there is still detectable flux in the presence of a wide
companion which should be represented by a point on the plot at the
wider projected separation.  For the pair with the closer projected
separation, there {\em may\/} be detectable submillimeter flux from a
close pair, and thus we adopt the conservative approach of putting a
point on the plot to represent this possibility.  Since we have argued
in Paper I that close pairs have smaller fluxes on average than wide
pairs, this method of assigning fluxes provides the most rigorous test
of our hypothesis.

Two triple systems, SR 24 and UZ Tau, have been treated
differently because the wide pair in each has been resolved at
millimeter wavelengths.  SR 24 N and S are separated by 6\arcsec, and
SR 24N is a close binary with a projected separation of 0\farcs2
(Simon et al.\ 1995).  AM mapped the area around SR 24 using an
``on-the-fly'' mapping technique and a 12\arcsec\ (FWHM) beam, and
they conclude that the detected 1300 \micron\ emission arises from an
unresolved area around SR 24S, with an upper limit of 30 mJy on
emission from SR 24N, the close binary.  They give fluxes of 280 mJy
for SR 24S and $<30$ mJy for SR 24N, though they note that the latter
limit is uncertain due to possible contamination from SR 24S\null.
This is consistent with the results of Reipurth et al.\ (1993), who
found the total SR 24 flux in a 23\arcsec\ beam to be $259 \pm 14$ mJy
at 1300 \micron.  We adopt the flux values given by AM for our
analysis.  For UZ Tau, the wide pair has been resolved at 1300
\micron\ in interferometric observations using the Owens Valley
Millimeter Array (Jensen, Koerner, \& Mathieu 1995, in preparation).
UZ Tau W, the close binary, has a flux of $32 \pm 9$ mJy at 1300 \micron,
while UZ Tau E has a flux of $137 \pm 28$ mJy.  In our analysis we have
assigned UZ Tau W its measured 1300 \micron\ flux and assigned the
combined flux of both components to the wide E-W pair.  Because of the
availability of observations which resolve the wide pairs, these two
systems are not marked as triples in Figures
\ref{fig:sco-oph}--\ref{fig:clearing-model}.

\subsection{The submillimeter flux distribution as a function of
  binary separation}

Figure \ref{fig:sco-oph} shows the 1300 \micron\ flux of Sco-Oph
binaries plotted as a function of projected binary separation.  Filled
symbols are detections and open symbols represent 3$\sigma$ upper
limits.  Squares show measurements at 1300 \micron, while triangles
are 800 \micron\ measurements scaled to 1300 \micron\ as described
above. To aid in determining which points may be confused by the
presence of higher-order multiplicity, each point representing a pair
in an unresolved higher-order multiple system is marked with a
``\multiplemark'' to the right of the point.  Different pairings from
the same system can be associated by having the same flux measurement.
(Note that in three instances the systems include a spectroscopic
binary with a separation smaller than the limit of the figure.)  All
fluxes have been scaled to a distance of 140 pc, roughly the mean
distance between the 125 pc distance to $\rho$ Oph and the 160 pc
distance to Upper Sco.

A dependence of submillimeter flux on projected binary separation is
evident in Figure \ref{fig:sco-oph}. Binaries with projected
separations $a_p$ less than roughly 100 AU tend to have lower fluxes
than wider binaries. Among binaries with $a_p \ge 100$ AU, 7 of 15 are
detected, while only 3 of 16 closer systems are detected, and two of
these have separations very near 100 AU\null. More importantly, the
close binaries which were not detected have flux upper limits
substantially lower than most of the detections among the wider
binaries. Clearly the submillimeter flux distribution differs between
binaries with separations greater or less than roughly 100 AU, with
the distribution extending to larger fluxes among the wider binaries.

In order to quantify the dependence of submillimeter flux on
binary separation, we applied statistical tests to compare the
distributions of flux among different groups of binaries.  We used the
techniques of survival analysis which allow quantitative analysis of
data which include upper limits.  We used the software package ASURV
Rev.\ 1.1 (LaValley, Isobe, \& Feigelson 1992) which implements the
methods presented in Feigelson \& Nelson (1985).  We divided the data
into subsets based on projected separation and then compared pairs of
these subsets using the Gehan, logrank, Peto-Peto, and Peto-Prentice
two-sample tests.  These tests yield probabilities that the
distributions of flux in the two samples are drawn from the same
parent distribution.  The various tests are sensitive to different
underlying flux distributions; since the true flux distribution for
our sample is unknown, we have applied all of the tests and report the
spread of the results.

We have divided the sample into close and wide binaries using
separations of both 50 AU and 100 AU as in Paper I\null. The results
of the two-sample tests are given in Table~\ref{table:asurv_results},
expressed as percentage confidence levels that the flux distributions
of the two samples are different.  These results support the
conclusion reached by visual examination of Figure \ref{fig:sco-oph}:
at the 90--95\% confidence level binaries with projected separations
less than or equal to 100 AU do not have the same submillimeter flux
distribution as binaries with greater projected separations.  When the
sample is divided at 50 AU, the same result is found with similar
confidence levels.

Simon et al.\ (1992, 1995) found that young binaries as a group (i.e.\
of all separations) have lower submillimeter fluxes on average than
young single stars.  We have also compared the sample of ``single''
stars to the known binaries, but we have considered the close and
wide binaries separately.  The confidence levels at which the
distributions of binaries and single stars differ are given in
Table~\ref{table:asurv_results}.  The close binaries have a different
flux distribution from the single stars at confidence levels around
95\%.  In contrast, the fluxes from the wide binaries and single
stars are consistent with being drawn from the same distribution.

We have applied the same two-sample tests discussed above to the PMS
binaries and single stars in the Taurus-Auriga star-forming region.
These results are also given in Table~\ref{table:asurv_results}, and
the Taurus-Auriga binaries are plotted in Figure \ref{fig:tau-aur}.
As expected, the results are the same as from similar previous
analyses (Paper I; Osterloh \& Beckwith 1995): the submillimeter flux
distribution of the close binaries is distinct from that of the wide
binaries at high confidence levels.  Thus, the same variation of
submillimeter flux with binary separation is seen in two independent
samples of PMS stars.  Finally, we combine the Sco-Oph and Tau-Aur
samples (Figure \ref{fig:combined-sample}), obtaining confidence
levels greater than 99\% that the submillimeter flux distributions of
close and wide binaries, and of close binaries and single stars, are
distinct.

Hence we conclude that at a statistically secure level the
submillimeter flux distribution among binaries with separations
between 1 AU and 50--100 AU is different than that of wider binaries
or single stars. The sense of the difference is that submillimeter
emission is weaker in close binaries.

The binary separation at which the transition occurs is not well
defined but appears to be roughly 50 AU to 100 AU\null.  This
separation is comparable to the radius typically derived for dust
disks around PMS stars (e.g.\ Beckwith \& Sargent 1993).  As pointed
out by Osterloh \& Beckwith (1995), this transition separation
strongly suggests that the emitting material does lie in disks in most
cases. If the material were in extended envelopes (e.g.\ Terebey et
al.\ 1993), binaries at this separation would not be expected to
influence the submillimeter emission.

The low submillimeter fluxes from close binaries are not due to age;
the closer binaries have the same distribution of ages as the wider
binaries or single stars.  In addition, within the range of ages
(roughly $10^5$--$10^7$ yr) in our sample, we find no dependence of
submillimeter flux on age of the system for the wide binaries.

\section{Interpreting the relationship between disk emission and
binary separation}

\subsection{Resonant clearing of gaps in disks}

The presence of a stellar companion at a separation less than the disk
radius must influence the spatial distribution of the disk material.
Theoretical calculations and numerical simulations show that a binary
embedded within a disk will rapidly clear a region on the size scale
of the binary separation, thus isolating circumstellar and
circumbinary disks (see e.g.\ Lin \& Papaloizou 1993 for a review).
The extent of the region cleared depends on the details of the given
system, specifically the binary mass ratio, orbital eccentricity, and
disk viscosity.  Artymowicz \& Lubow (1994) find that circumstellar
disks will have outer radii of less than half the binary semi-major
axis, and circumbinary disks will have inner radii of greater than
roughly twice the binary semi-major axis.

Disk clearing of this type creates a natural link between binary
separation and the submillimeter fluxes from binary systems which is
very similar to that observed.  Assume that disks have a certain
characteristic outer radius $R_d$.  For the purposes of this
discussion, this radius could be either a physical limit on the extent
of disk material or an effective radius inside which most of the
disk's submillimeter emission originates (even though there may be
additional cold or low-density material outside it).  In binary
systems with separations greater than a few times $R_d$, each
component of the system could have a circumstellar disk that is
relatively undisturbed by its companion, and the submillimeter
emission from such a system would be comparable to that from a single
star with a circumstellar disk.  In systems with separations much less
than $R_d$ only a small hole at the center of the circumbinary disk
will be cleared, leaving a circumbinary disk which could be
similar in extent and emission to disks around single stars.  Binaries
with separations somewhat less than $R_d$, however, will
clear regions whose extent represents a large fraction of the surface
area of an undisturbed disk.  This reduction in emitting surface area
will result in reduced submillimeter fluxes, as observed.

In order to investigate the effect of gap clearing on submillimeter
flux in a more quantitative way, we have introduced gaps into a simple
disk model.  Following a standard approach in modeling infrared and
submillimeter disk emission (e.g.\ Adams, Lada, \& Shu 1988, Adams,
Emerson, \& Fuller 1990, BSCG), we use a geometrically thin disk with
power-law temperature and surface-density distributions:
\begin{equation}\label{eq:temperature}
  T(r) = T_1(r/{\rm (1\ AU)})^{-q},
\end{equation}
\begin{equation}
  \Sigma(r) = \Sigma_0(r/r_0)^{-p}.
\end{equation}
The emission from such a disk is then given by
\begin{equation}\label{eq:disk_emission}
  F_{\nu} = \frac{\cos\theta}{D^2} \int_{r_0}^{R_d} B_{\nu}[T(r)] (1 -
  e^{-\tau_\nu(r)})2\pi r \, dr,
\end{equation}
where $D$ is the distance from the observer to the system, $\theta$ is
the angle between the line of sight and the normal to the disk plane,
and $\tau_\nu$ is the line-of-sight optical depth through the disk,
related to the mass opacity $\kappa_\nu$ by
\begin{equation}
  \tau_\nu(r) = \frac{\kappa_\nu \Sigma(r)}{\cos \theta}.
\end{equation}

To simulate disk clearing by a binary companion, we made a
simple modification to this disk model.  For each model binary-disk
system, we set the disk surface density (and thus the emission) equal
to zero between radii $r_{in}$ and $r_{out}$, designating the inner
and outer edges of the gap:
\begin{equation}\label{eq:surface_density}
  \Sigma(r) = \cases{0,& if $r_{in} < r < r_{out}$;\cr
              \Sigma_0(r/r_0)^{-p},& otherwise.}
\end{equation}
These radii are determined by the parameters of a given binary system,
especially the binary semi-major axis $a$.  We considered two
different cases.  In the first, $r_{in} = 0.4a$ and $r_{out} = 1.8a$.
This is the clearing expected for a binary with a circular orbit, and
the minimum clearing expected for any binary system (Artymowicz \&
Lubow 1994).  In the second case, $r_{in} = 0.2 a$ and $r_{out} = 3a$.
This level of clearing is expected for a binary with an orbital
eccentricity of 0.4, roughly the mean eccentricity for \pms\ and
main-sequence binaries with known orbits (Mathieu 1994, Duquennoy \&
Mayor 1991).\footnote{These clearing radii also depend on the disk
  viscosity and the binary mass ratio.  The numbers quoted here are
  for a Reynolds number of $10^5$ and a binary mass ratio of 7:3.
  However, changing these values within reasonably expected bounds
  does not greatly affect the expected disk clearing.  For a range of
  Reynolds numbers from $10^4$ to $10^6$ and eccentricity of 0.4, the
  inner gap edge ranges from $0.19a$ to $0.25a$ and the outer gap edge
  from $2.7a$ to $3.1a$ (Artymowicz \& Lubow 1994).  We adopt the
  numbers above as representative of an ``average'' binary.}

For a given set of disk parameters ($M_d$, $q$, $T_1$; $p = 1.5$, $R_d
= 100$ AU) we calculated the expected submillimeter flux from binaries
having a range of separations from 0.01 AU to 2000 AU, spanning the
observed range of separations in our sample.  We then calculated
models with a range of inclination angles and averaged their fluxes
based on the frequency of occurrence of a given value of $\theta$ in a
random distribution of disk inclinations.  The result for one
particular disk model is shown in Figure \ref{fig:clearing-model}.
The model is superimposed on the combined submillimeter data from the
Sco-Oph and Tau-Aur regions, with all fluxes scaled to a distance of
140 pc.  The solid line shows the expected submillimeter flux as a
function of separation if gaps extend from $0.2 a$ to $3 a$ (eccentric
orbits), and the dashed line shows the expected flux if gaps extend
from $0.4 a$ to $1.8 a$ (circular orbits).

As expected, the gap-clearing model shows the submillimeter fluxes of
the closest and widest binaries to be unaffected by gap clearing,
while binaries of intermediate separations have lower fluxes. For the
particular model shown the reduction in flux occurs for binaries with
separations between roughly 1 AU and 300 AU, with a minimum flux
around 25 AU\null.  Thus our simple model for gap clearing reproduces
the qualitative features of the observed submillimeter flux dependence
on binary separation.

If the gaps are of the size expected for eccentric binary systems, the
reduction of submillimeter flux due to gaps can be as much as a factor
of 15. This is comparable to the difference between typical detected
fluxes among wide binaries and the available upper limits for the
close binaries.  The model shown in Figure \ref{fig:clearing-model}
provides a specific case for quantitative discussion. This model has
$M_d = 0.05\ \msun$, $q = 0.65$, and $T_1 = 125\ {\rm K}$ and was
chosen to match typical flux levels among detected wide binaries.  The
minimum submillimeter flux in this particular model is a factor of
nine less than the flux of the disk with no gap. The dashed line
(circular binary) lies well above most of the observed flux upper
limits between 1 and 100 AU and thus is not consistent with the
observations.  However, most binaries have eccentric orbits.  The
solid line ($e=0.4$) passes through the upper limits for
binaries with separations around 25 AU and exceeds the typical upper
limits for binaries of somewhat larger and smaller separations by a
factor of two to three at most.

{\em The essential conclusion to be drawn from this simple model is
  that the lack of detected submillimeter emission among the closer
  binaries does not necessarily imply that disk material is not
  present, nor that disk surface densities are much lower than found
  among wide binaries or single stars. The large reduction in emitting
  surface area due to gap clearing can in and of itself reduce flux to
  levels comparable to many of the present upper limits.} Thus, while
it is clear from these submillimeter data that binary companions with
separations less than 100 AU significantly influence the nature of
associated disks, it remains an outstanding question to what extent
circumstellar and circumbinary disks are present in these binaries. We
return to these issues in Sections \ref{sec:cs-disks} and
\ref{sec:cb-disks}

Given the presence of a companion, the physical conditions of the
disk(s) in our simple model are certainly oversimplified, and thus a
detailed comparison of the model with observations is likely not
merited. Nonetheless, we note that the model appears less successful
for binaries with separations between 1 AU and 10 AU\null. In
particular, the model flux is symmetric in $\log(a_p)$ around binary
separations of a few tens of AU, while in fact submillimeter fluxes
remain low at separations less than 10 AU\null.  Unfortunately, the
sample size in this separation range is small; the two-sample tests
give probabilities of only 78--90\% that the submillimeter fluxes of
binaries with $1 < a_p < 10$ AU are drawn from a different
distribution than the fluxes of binaries with $a_p > 100$ AU\null.
Furthermore, most of the binaries with $1 < a_p < 10$ AU were
discovered via lunar occultation techniques, so that the measured
separation is a projection against the lunar limb.  As such, many of
these binaries may have wider separations projected on the sky than
those shown in Figure \ref{fig:clearing-model}.  Thus, while there is
a suggestion that there is a larger reduction in flux among the
closest binaries than predicted by the model, the case is not strong
with the present data.

If the reduction in submillimeter flux does extend to separations as
small as 1 AU, this would be difficult to reproduce with our simple
annular gap model. We note that the flux predicted by the model for
binaries with separations of less than 10 AU is largely from
circumbinary disks. Indeed for binaries with separations of only a few
AU the surface areas of circumstellar disks are too small to produce
detectable submillimeter emission.  Hence weak submillimeter emission
from such binaries would suggest lower mass, temperature, and/or dust
opacity of circumbinary material.

Finally, we note that the most submillimeter-luminous binaries, such
as GG Tau, T Tau, and AS 205, cannot be easily incorporated into this
simple picture.  GG Tau in particular has a projected separation of 40
AU and yet is the most luminous binary in the Tau-Aur and Sco-Oph
regions.  Furthermore, interferometric observations at millimeter
wavelengths reveal a circumbinary disk with a radial extent of at
least 800 AU (Dutrey, Guilloteau, \& Simon 1994). The millimeter
emission also shows a central depression which Dutrey et al.\
attribute to a central cavity of radius 180 AU\null. They attribute this
cavity to a clearing process similar to that invoked here for gaps,
but clearly this system is substantially different from both our
standard model for a 40 AU binary and from other binaries with
projected separations of tens of AU\null.

\subsubsection{Effect of disk parameters on the flux-separation relation}
\label{sec:disk-params}

The model shown in Figure \ref{fig:clearing-model} is only one
realization of a large range of physical conditions found in disks.
Therefore it is useful to establish whether other combinations of
model parameters can better reproduce the observed variation of flux
with separation.  In Figure \ref{fig:4-clearing-models} we explore the
influence of disk parameters on the distribution of submillimeter flux
with binary separation.  In panels a, b, and c, one of the disk
parameters $T_1$, $q$, and $M_d$ is varied while the others are held
fixed.  In panel d, all three disk parameters are held fixed while the
wavelength at which the flux is calculated is varied.

One general trend that can be seen in these figures is that changes in
disk properties which give lower submillimeter fluxes also tend to
decrease the binary separation corresponding to the minimum flux.
This is because the contribution of the hotter, more optically thick
inner regions of the disk contribute a larger percentage of the total
flux when the disk luminosity is low.  For more luminous disks whose
outer regions are hotter (through higher $T_1$ or lower $q$) or more
optically thick (through higher $M_d$), the contribution of the outer
disk to the total flux is significant. In such disks gap clearing
in the outer regions has a larger effect on the total submillimeter
flux.

A discontinuous change in slope can be seen in some of the models in
Figure \ref{fig:4-clearing-models} (as well as in Figure
\ref{fig:clearing-model}).  This is due to the assumption in the
models that the disks have a sharp outer radius $R_d$, here taken to
be 100 AU\null.  The slope discontinuity occurs when the outer edge of
the gap $r_{out}$ reaches the outer edge of the disk.  Increasing
$R_d$ would increase the separation that gives the minimum flux for a
few of the models shown in Figure \ref{fig:4-clearing-models}, though
most would be unchanged.

Interestingly, the choice of 1300 \micron\ or 800 \micron\ as an
observing wavelength makes little difference in the
binary separation producing the minimum flux (Figure
\ref{fig:4-clearing-models}d).  This is due to the interplay of two
opposing factors.  Shorter-wavelength emission is more sensitive to
disk material at higher temperatures, suggesting that 800 \micron\
emission would tend to probe gaps at smaller disk radii than 1300
\micron\ emission.  However, for emission in the Rayleigh limit the
tradeoff between increasing surface area and decreasing optical depth
with radius tends to emphasize the emission from the region in the
disk near $\tau_\nu = 1$ (BSCG).  Because dust opacity increases with
frequency, $\tau_\nu = 1$ lies at a larger radius for 800 \micron\
than for 1300 \micron.  These two effects largely cancel each other
and emission at both wavelengths is sensitive to roughly the same disk
radii.  However, 60 \micron\ flux is sensitive to smaller disk radii
because much of the disk is too cold to produce appreciable 60
\micron\ emission.

We draw two main conclusions from this exploration of model parameter
space.  First, under no circumstances does gap clearing at the level
expected for binaries with circular orbits produce a reduction in flux
that is comparable to that observed.  Second, no combination of model
parameters does significantly better than that in Figure
\ref{fig:clearing-model} at reproducing the depth, breadth, and
location of the reduction in submillimeter fluxes observed among the
close binaries.

\subsubsection{Estimates of disk masses}\label{sec:masscalc}

Gap clearing may affect not only the distribution of submillimeter
flux with separation, but also the disk mass derived for a given
system based on its submillimeter flux.  Submillimeter fluxes have
commonly been used to infer disk masses by adopting a model for a
continuous disk, deriving the disk temperature distribution from
infrared data, and then adjusting the disk mass until the model flux
agrees with the observed submillimeter flux at one or more wavelengths
(Adams et al.\ 1990, BSCG, AM, Osterloh \& Beckwith 1995).  However,
if binaries clear gaps in their disks, the geometry of the emitting
material is significantly different than that assumed in conventional
disk modeling and may affect the mass calculation.  Here we derive
disk masses assuming the disk-clearing model discussed above and
compare them to masses derived assuming continuous disks.  Because
disk clearing has little effect on disk mass estimates for wider
binaries, we derive masses only for Sco-Oph binaries with $a_p < 300$
AU\null.

The temperature distribution of the disk is determined by fitting the
10--100 \micron\ data for each system with a power law and determining
$q$ and $T_1$ from the parameters of the fit (BSCG).  Some of the
binaries in our sample do not have sufficient infrared data available
in the literature for this to be possible, and we could not derive
masses for these systems.  For the remaining systems, we took published
values of $q$ and $T_1$ from AM for some sources and used infrared
data from the literature to derive $q$ and $T_1$ for the others.
Though emission at 10--100 \micron\ is likely to be dominated by the
inner part of the circumstellar disk, we assume that the derived
temperature distribution applies to the whole disk.

To calculate the disk mass in the presence of a gap, we assume the
emission from the disk is given by Eq.\ \ref{eq:disk_emission}, with
the disk surface density given by Eq.\ \ref{eq:surface_density}.  We
calculate masses for three cases: no clearing, circular-orbit clearing
($r_{in} = 0.4a$, $r_{out} = 1.8a$), and eccentric-orbit clearing
($r_{in} = 0.2a$, $r_{out} = 3a$).  For triple or quadruple systems,
we choose the projected separation closest to 50 AU since this pair
will most affect the derived disk mass.  Following BSCG, we take $p =
1.5$, $\theta = 0$, $r_0 = 0.01$ AU, and $R_d = 100$ AU\null.
Following Beckwith \& Sargent (1991), we use the opacity law
$\kappa_\nu = 0.1\,(\lambda/250\ \micron)^{-\beta}$, with $\beta = 1$.
We then numerically integrate Eq.\ \ref{eq:disk_emission} and vary
$\Sigma_0$ until the calculated flux matches the observed flux.  These
masses are given in Table \ref{table:disk_masses}.

The table shows that gap clearing does not greatly affect the derived
mass for most systems.  We further explored this affect using an
ensemble of model binary systems with a range of disk properties.  For
gaps from $0.4 a$ to $1.8 a$, the derived disk mass typically differs
by a factor of two or less from the mass calculated assuming a
continuous disk; for gaps from $0.2 a$ to $3 a$, the difference is
typically a factor of three or less.  For some combinations of disk
properties and gap locations, the variations are somewhat greater.
Gap clearing can either increase or decrease the mass derived from a
given flux, depending on whether the area cleared is efficient or
inefficient at radiating submillimeter flux compared to the rest of
the disk.  Thus, gap clearing introduces an additional uncertainty
into disk mass calculations for close binary systems.

\subsection{Circumstellar disks in binary environments}\label{sec:cs-disks}

While gap clearing can plausibly explain the reduction of
submillimeter flux from binaries
with separations of less than 50-100 AU, the upper limits on
fluxes from such binaries can be equally well explained if the physical
conditions (such as surface densities or temperatures) of disks in
closer binaries differ from those in wide binaries. Indeed, one
straightforward interpretation of the lack of submillimeter emission is
the complete absence of disk material.

However, 2 \micron\ and 10 \micron\ excesses indicate that
circumstellar disks are present in at least 50\% of PMS binaries
(Mathieu 1994).  Similarly, Simon et al.\ (1995) find the same binary
frequency in systems with and without circumstellar disks based on $K
- L$ colors, and Simon \& Prato (1995) find the same frequency of $K -
N$ color excesses for single stars as for binaries.  But
observations at these wavelengths only sample material very near
stellar surfaces.  In addition the high dust opacities at these
wavelengths typically do not permit derivation of surface densities.

Longer wavelength IRAS measurements also support a high frequency of
circumstellar disks.  Mid-infrared (e.g.\ 60 \micron) flux originates
in the inner regions of disks (typically $\le 10$ AU); thus for most
of the binaries in our sample 60 \micron\ flux originates in
circumstellar disks.  Figure \ref{fig:iras} shows IRAS 60 \micron\
flux plotted as a function of projected separation for the Tau-Aur and
Sco-Oph binaries.  IRAS fluxes or upper limits were taken from Weaver
\& Jones (1992), Strom et al.\ (1989), Clark (1991), Hartmann et al.\
(1991), Wilking, Lada, \& Young (1989), and the IRAS Point Source
Catalog.  Filled symbols represent detections and open symbols are
upper limits.  All fluxes are scaled to a distance of 140 pc.
Fourteen systems out of 85 do not have IRAS measurements in any of the
above references and are not shown here.  These systems are
approximately equally divided between wide and close binaries.

Figure \ref{fig:iras} has two notable features.  First, the fraction
of binaries detected at 60 \micron\ is much higher than at
submillimeter wavelengths.  If 60 \micron\ flux is taken to originate
in disks, then at least one circumstellar disk is present in most PMS
binaries.  As such, the low level of submillimeter emission from the
close binaries is not the result of a total absence of disk material.
Second, there is no marked dependence of 60 \micron\ flux on binary
separation akin to that seen at submillimeter wavelengths. The same
two-sample tests performed on the submillimeter data show no
difference in the 60 \micron\ flux distributions of the close and wide
binaries divided at 10, 50, or 100 AU\null.

These 60 \micron\ flux measurements can provide meaningful constraints
on the surface-density normalizations of circumstellar disks. Because
dust opacity is much higher at 60 \micron\ than at submillimeter
wavelengths, disks remain optically thick at 60 \micron\ to much lower
surface densities than in the submillimeter. An explicit assumption of
previous analyses of disk masses is that the disks are optically thick
at 60 \micron\ and thus reflect the disk temperature distributions.
However, given only an upper limit on submillimeter flux and thus
circumstellar disk surface density, this assumption need not hold.
For a power-law radial temperature distribution one signature of
optically thin 60 \micron\ emission would be a steepening of the
spectral slope between 12 \micron\ and 60 \micron.  In fact, several
binaries show such steepening at a formally significant level given
the quoted uncertainties on the IRAS fluxes.  Thus some circumstellar
disks may be partially optically thin at 60 \micron.  However, given
that other binaries also show deviations from single spectral slopes
in other senses (e.g., high 12 \micron, 25 \micron, or 60 \micron\
fluxes at formally significant levels), we do not feel that in any
given case the conclusion of optical thinness is secure.

Rather we choose to use the binaries with steepening spectral slopes
to derive lower limits on masses and surface densities of
circumstellar disks.  Specifically we have chosen three Tau-Aur
binaries with $a_p < 50$ AU and for which the slope of $\lambda
F_\lambda$ from 25 to 60 \micron\ is steeper than the slope from 12 to
25 \micron: DF Tau, FO Tau, and CZ Tau. These binaries have the
highest quality IRAS flux measurements and no additional
companions.\footnote{CZ Tau lies 30\arcsec\ from DD Tau, but Weaver \&
Jones (1992) list CZ Tau as a ``better positional fit'' to the IRAS
source.}  The IRAS fluxes of these three binaries lie in the middle of
the range of detected fluxes, and thus we take these systems to be
representative examples of young binary systems.  For each binary we
have derived disk masses using the method described in \S
\ref{sec:masscalc}, except that the disk temperature was derived from
only the IRAS 12 and 25 \micron\ fluxes and the disk radius was taken
to be half the projected binary separation.  The disk surface density
was then varied so that the model flux matched the observed 60
\micron\ flux.  We used the dust opacity law given by Adams et al.\
(1988) since Beckwith \& Sargent (1991) do not give opacities for IRAS
wavelengths.

The circumstellar disk masses derived for these binaries range from $5
\times 10^{-6}$ \msun\ to $7 \times 10^{-5}$ \msun. These numbers
cannot be directly compared with masses derived for disks around
single stars because of the smaller disk radii used here.  A more
significant comparison is the disk surface-density normalization
($\Sigma_0$; Eq.\ \ref{eq:surface_density}).  The derived
surface-density normalizations for these circumstellar disks are roughly two
orders of magnitude smaller than those for disks around single stars
detected at submillimeter wavelengths.  {\em The essential conclusion
  to be drawn from this analysis is that the mechanism reducing the
  submillimeter flux from binaries with $1 < a_p < $ 50--100 AU does
  not entirely destroy circumstellar disks or inhibit their
  formation.}

We stress that these mass and surface density estimates are best
considered as lower limits.  While the steepening spectral slopes of
these binaries may be indicative of partial optical thinness at 60
\micron\ as presumed in these calculations, we cannot rule out
fluctuations in the IRAS fluxes larger than the formal errors.  At the
same time, we have shown that in the presence of gaps the
submillimeter flux upper limits do not require that the circumstellar
surface densities be significantly lower than found in disks around
single stars.  Thus, the available data require circumstellar disk
surface densities in close binaries to be in a range from 1\% to the
same as typical disks around single stars.  These limits constrain the
degree of depletion of circumstellar disks by the various processes
discussed in the Introduction.

The lack of dependence of 60 \micron\ emission on binary separation is
not inconsistent with the gap model. The specific model shown in Figure
\ref{fig:4-clearing-models}d predicts a factor of $\sim 5$ decrease in
60 \micron\ flux for binaries with separations between 1 AU and 10
AU\null.  Given the small number of systems in this separation range,
such a change would be undetectable with current data.  In addition,
as noted previously most binaries at these small separations were
discovered via lunar occultation, so that their true separations may
be substantially underestimated.

Finally, it is plausible that the temperature distributions of disks
in close binary environments would differ from those in wide binaries
or around single stars, also influencing submillimeter and infrared
emission.  We used the $T_1$ and $q$ values (see Eq.\
\ref{eq:temperature}) for the Sco-Oph and Tau-Aur binaries to
investigate whether the temperature parameters vary systematically
with separation.  We find no evidence for a dependence of disk
temperature on binary separation. These parameters only provide
information about the inner 10 AU or so of the disk. The temperature
distribution of the disks elsewhere---and in particular at radii
comparable to most binary separations---is largely unconstrained by
current observations.

\subsection{Circumbinary disks}
\label{sec:cb-disks}

Because of their low temperatures, circumbinary disks in all but the
closest binaries emit predominantly at far-infrared and longer
wavelengths. However, since substantial emission at these wavelengths
can arise from a large range of radii within a disk, securely
associating unresolved flux measurements with circumbinary disks can
be problematic. On the other hand, upper limits on submillimeter
emission unambiguously place upper limits on circumbinary disk masses.
For optically-thin emission at $\lambda = 1300$ \micron\ with
$\kappa_\nu = 0.02$ cm$^{2}$ g$^{-1}$, the circumbinary mass $M_{cb}$
that produces a given flux $F_\nu$ is
\begin{equation}\label{eq:cb-mass}
  M_{cb} = 2.6 \times 10^{-4}\, \msun\ (F_\nu/{\rm mJy})\
  (e^{(11.1\ {\rm K}/T)} - 1)
\end{equation}
assuming a distance of 140 pc and emission in the Rayleigh limit.
Taking $T=15$ K as a minimum dust temperature and $F_\nu < 15$ mJy as
a typical flux upper limit for binaries with separations of less than
$\sim 100$ AU, we find $M_{cb} < 4.3 \times 10^{-3}$ \msun. This upper limit is
conservative given that some emission may derive from a circumstellar
disk. On the other hand if the circumbinary material were optically
thick, the derived mass limit would be higher.  {\em Thus we conclude
  that binaries with projected separations between a few AU and $\sim 100$
  AU typically do not have circumbinary disks with masses greater than
  0.005 \msun.}

There is one notable counterexample, however. As discussed above, a
ring-like circumbinary disk around the 40 AU binary GG Tau has been
resolved at 3 mm (Dutrey et al.\ 1994).  The mass derived for this
ring from the continuum emission is 0.13 \msun. GG Tau is unusual in
being the most luminous submillimeter source in our sample (Figure
\ref{fig:combined-sample}) and one of the younger binaries in Tau-Aur.
Most PMS binaries of similar projected separation do not have
similarly massive or luminous circumbinary disks, suggesting the
possibility that GG Tau represents a brief phase in early circumbinary
disk evolution.

Circumbinary disks have also been found around several very close
binaries.  For example, in HP Tau (one of the closest binaries in our
sample with detected 1300 \micron\ emission), a simple calculation
shows that a circumbinary disk must be present.  If the binary orbit
were filled with optically-thick material emitting at the stellar
temperature, the resulting submillimeter flux would be comparable to
that observed; however, the optical flux of such a quantity of hot
material would be $\sim 10^4$ times the observed flux.  Thus, the
submillimeter-emitting region in HP Tau must be substantially larger
than a few AU in radius and therefore must lie primarily outside the
binary orbit.  A similar calculation requires the presence of
circumbinary disks around the PMS spectroscopic binaries GW Ori
(Mathieu et al.\ 1995), AK Sco and V4046 Sgr (detected in this work), and
DQ Tau (Mathieu 1995).

To conclude, our flux limits place upper limits on circumbinary disk
masses of 0.005 \msun\ for most binaries with separations of less than
roughly 100 AU\null. Indeed, it remains possible that most binaries do
not have circumbinary disks. The present submillimeter flux detections
and upper limits for almost all binaries with separations greater than
a few AU are entirely consistent with emission from only circumstellar
disks.

\section{Conclusion}
We have made sensitive 800 \micron\ continuum observations of most
(25) \pms\ binaries with projected separations $a_p \lesssim 150$ AU
in the Scorpius-Ophiuchus star-forming region, and we have
supplemented these data with previous 1300 \micron\ continuum
observations to obtain a sample of \numophbinaries\ systems. We have
also created a similar database from the literature for
\numtaubinaries\ systems in Taurus-Auriga. We have used these data to
study the nature of disks in young binary environments, and we find:

1) Submillimeter fluxes from binaries with $1 < a_p <$ 50--100 AU are
lower on average than from wider binaries or single stars, whereas the
flux distributions of wide binaries and single stars are
indistinguishable.  This dependence of submillimeter flux on binary
separation is seen independently in the Sco-Oph and Tau-Aur samples.
When the samples are combined the effect is found at greater than the
99\% confidence level. The transition separation of 50--100 AU is
similar to typically derived radii for dust disks around young stars,
strongly suggesting that the reduction in submillimeter emission among
closer binaries is due to the influence of the companions on disks.

2) The reduction in submillimeter flux from binaries with $1 < a_p <$
50--100 AU can plausibly be attributed to gaps cleared in disks by
binaries with eccentric orbits. As such, the present upper limits
permit but do not require a large reduction in the surface densities
of disk material outside such gaps compared to surface densities of
disks among wide binaries or single stars.

3) Most of the binaries in our sample were detected at 60 \micron\ by
IRAS, indicating that each of these binaries has at least one circumstellar
disk. Presuming the disks to be optically thick at 60 \micron, the
flux measurements place lower limits of roughly $10^{-5}$ \msun\ on
circumstellar disk masses.  This lower limit corresponds to circumstellar
disk surface densities no more than two orders of magnitude smaller
than surface densities of most disks detected at submillimeter
wavelengths.

4) We place upper limits of 0.005 \msun\ on circumbinary disk masses
around most binaries with projected separations between 1 AU and 100 AU\null.
Circumbinary disks as massive as that found around GG Tau
are rare for these binaries.  However,
submillimeter detections of binaries with
separations less than a few AU show that massive circumbinary disks
can exist around the closest binaries.

The present body of data is consistent with the following picture for
disks in PMS binary environments.  Binaries with semimajor axes of a
few hundred AU or greater have circumstellar disks with
properties very similar to those of disks found around single stars.
Binaries with semimajor axes between a few AU and 50--100 AU also
typically have at least one circumstellar disk each. The binaries truncate
these disks, thus limiting their submillimeter emission. Their surface
densities remain uncertain within a range of 0.01 to 1 times the
surface densities of disks around single stars. Binaries in this
separation range typically do not have massive ($M > 0.005$ \msun)
circumbinary disks.  Binaries with semimajor axes of less than a few
AU can have massive circumbinary disks. Such binaries truncate the
inner edges of circumbinary disks at radii of only a few AU or less,
leaving most of the disk undisturbed on dynamical timescales.

Notably absent from this morphological picture is any discussion of
the detailed physical conditions of the disks, such as temperature,
surface-density, and opacity distributions in both circumstellar and
circumbinary disks. In addition to clearing gaps, companions are
expected to influence these disk properties (e.g.\ Syer \& Clarke
1995).  At the same time, the present submillimeter flux upper limits
can only just be explained by our model, even assuming relatively
large gaps driven by eccentric binaries. We anticipate that disks in
young binary environments are substantially more complex than the
simple disk model employed here.

Finally, we note that binaries are the primary product of star
formation, and thus the frequency of other planetary systems is
intimately linked to the issues raised here. For binaries with
separations of less than 100 AU, present upper limits on submillimeter
flux imply upper limits on total disk mass of a few times $10^{-3}$
\msun. This is roughly an order of magnitude smaller than typical
estimates for the minimum mass of the early solar nebula (e.g.\
Weidenschilling 1977). Such a disk could conceivably form terrestrial
planets, but it is unlikely to form gaseous giant planets.  However,
these mass estimates are insensitive to material in grains or
planetesimals larger than a few mm in size.  Thus, it remains possible
that close binaries can form planets, but it would appear that
sufficient disk material to form planetary systems like our own is
most likely to be found in wide binaries or single stars.

\acknowledgements{ We thank the referee, Steve Beckwith, for useful
  comments which improved the presentation of this paper.  We are
  grateful to the staff of the JCMT for their knowledgeable support
  and good humor. ELNJ gratefully acknowledges the support of a
  Grant-in-Aid of Research from the National Academy of Sciences
  through Sigma Xi, as well as funding from the National Space Grant
  College and Fellowship Program and the Wisconsin Space Grant
  Consortium.  RDM appreciates funding from the Presidential Young
  Investigator program, a Guggenheim Fellowship, the Morrison Fund of
  Lick Observatory, and the Wisconsin Alumni Research Fund.  GAF
  acknowledges the support of an NRAO Jansky Fellowship.  This
  research has made use of the Simbad database, operated at CDS,
  Strasbourg, France. }

\appendix

\section{Appendix: Observations of
  pre--main-sequence stars in other regions}
\label{sec:appendix}

During the course of the JCMT observations reported above, we surveyed
a number of other PMS binaries at submillimeter wavelengths.  These
systems were excluded from the main sample discussed above because
they are located in star-forming regions other than Sco-Oph or
Tau-Aur, or because they have spectral types earlier than F0.
Submillimeter fluxes for these systems are given in Table
\ref{table:other_data} and each is discussed briefly below.  While we
did not make a systematic survey, our detections of 800 \micron\
emission from the isolated T Tauri stars Hen 3-600, HD 98800, and
V4046 Sgr are interesting.  If these systems were formed from small,
isolated molecular clouds, our detections suggest that disk survival
times (at least in binary systems) may be longer than the cloud
dispersal times.

\hangpar {\bf V4046 Sgr}: V4046 Sgr (HDE 319139, HBC 662) was
discovered to be a spectroscopic binary by Byrne (1986) and de la Reza
et al.\ (1986). It is the shortest-period PMS binary detected at
submillimeter wavelengths to date and is also one of the brightest PMS
binary sources at these wavelengths. The strength of the submillimeter
emission requires the presence of a circumbinary disk (see
\S\ref{sec:cb-disks}).  V4046 Sgr is not associated with any known
region of star formation; it lies $1\deg$ from the nearest (small)
dark cloud and is not near any CO clouds (de la Reza et al.\ 1986).
However, it does show characteristics typical of classical T Tauri
stars: strong, variable H$\alpha$ emission, Li absorption, and UV and
infrared excesses (de la Reza et al.\ 1986).  V4046 Sgr has a disk
mass comparable to those found around other T Tauri stars, another
piece of evidence that it is indeed a \pms\ binary.

\hangpar
{\bf CoD $-$29\deg8887, CoD $-$33\deg7795, Hen 3-600}: These stars are
isolated T Tauri stars located in the vicinity of TW Hya (de la Reza
et al.\ 1989, Gregorio-Hetem et al.\ 1992).  All have H$\alpha$
emission and Li absorption, but they are not located near any known
star-forming region and are at moderately high (21\deg--28\deg)
galactic latitude.  Hen 3-600, the only one of the three detected at
800 \micron, is also the only one detected by IRAS.  Both
CoD $-$29\deg8887 and Hen 3-600 are binaries (de la Reza et al.\ 1989,
Reipurth \& Zinnecker 1993).

\hangpar
{\bf HD 98800}: This quadruple system (Torres et al.\ 1995) is another
isolated \pms\ system in the vicinity of TW Hya.  Zuckerman
\& Becklin (1993) noted that its infrared excess is large for a
Vega-excess system; however, the excess is not unusual for a PMS
system.  Fekel \& Bopp (1993) classify it as \pms\ with an
estimated age of less than $10^7$ yr.  The 800 \micron\ flux measured by
us is consistent with that reported by Rucinski (1993).  Our 800
\micron\ flux measurement, combined with the 1100 \micron\ measurement
by Stern, Weintraub, \& Festou (1994), yields disk masses of
$\sim 10^{-3}$--$10^{-4}$ \msun\ for assumed distances of 20--60 pc.

\hangpar
{\bf BZ Sgr}: This 5\arcsec\ binary was discovered to be a
high-latitude T Tauri system by Gregorio-Hetem et al.\ (1992).  Our
detection of 800 \micron\ emission is additional evidence that it is a
PMS system.

\hangpar
{\bf FK Ser}: A 1\farcs33 PMS binary (Herbig \& Bell 1988),
this system was classified as ``post-T Tauri'' by Herbig (1973) but
has since been considered a normal T Tauri star in many references
(e.g.\ Herbig \& Bell 1988).  Our detection of 800 \micron\ emission
further suggests that FK Ser has disk properties similar to those of
other T Tauri stars.

\hangpar
{\bf S CrA}: This 1\farcs4 PMS binary has a spectral type of
K6 and lies in the Corona Australis star-forming region (Herbig \&
Bell 1988).

\hangpar
{\bf HD 150193 (MWC 863, Elias 2-49)}: This Herbig Ae/Be binary has a
spectral type of A0 (Elias 1978) and a projected separation of
1\farcs1 (Reipurth \& Zinnecker 1993).

\hangpar
{\bf CoKu Ser/G1}: This 3\farcs5 PMS binary has a spectral
type of K7 and lies in the Serpens molecular cloud (Herbig \& Bell
1988).

\begin{table}[t]
\caption{\sc JCMT fluxes or $3\sigma$ upper limits for Sco-Oph binaries
\label{table:jcmt_data}}
\vspace*{4mm}
\begin{tabular}{lccccc}
\hline
Name & Projected & $F_\nu\ (1100\ \micron)$ & $F_\nu\ (800\ \micron)$ &
                   $F_\nu\ (450\ \micron)$ & $F_\nu\ (350\ \micron)$ \\
     & Sep\null. (AU)\tablenotemark{a} & (mJy) & (mJy) & (mJy) & (mJy) \\
\hline
\hline\\[-5mm]
155203$-$2338       & 128 &  \nodata &  $ <    30 $ &  \nodata &  \nodata \\
155808$-$2219       & 0.048 &  \nodata &  $ <    26 $ &  \nodata &  \nodata \\
155913$-$2233\tablenotemark{b}       & 0.014/46.1 &  \nodata &  $ <    26 $ &
\nodata &  \nodata \\
160814$-$1857       & 0.19 &  \nodata &  $ <    36 $ &  \nodata &  \nodata \\
160905$-$1859       & 0.015 &  \nodata &  $ <    34 $ &  \nodata &  \nodata \\
160946$-$1851       & 33.3 &  \nodata &  $ <    38 $ &  \nodata &  \nodata \\
162218$-$2420       & 29.5 &  \nodata &  $ <    39 $ &  \nodata &  \nodata \\
162814$-$2427\tablenotemark{b}       & 0.27/19.4 &  \nodata &  $ <    16 $ &
\nodata &  \nodata \\
162819$-$2423\tablenotemark{b}       & 0.10/2/600 &  $ <    30 $ &  $ <    17 $
&  \nodata &  \nodata \\
AK Sco            & 0.143 &  $    36 \pm  4 $ &  $    83 \pm  9 $ &  $   242
\pm 32 $ &  \nodata \\
AS 205            & 211 &  $   503 \pm 17 $ &  $   998 \pm 43 $ &  $  2541 \pm
190 $ &  $  3515 \pm 446 $ \\
DoAr 24E          & 257 &  \nodata &  $    37 \pm  8 $ &  \nodata &  \nodata \\
Elias 2-26        & 150 &  \nodata &  $ <    58 $ &  \nodata &  \nodata \\
Haro 1-4          & 90.0 &  \nodata &  $    53 \pm  9 $ &  $ <   361 $ &
\nodata \\
ROX 31            & 60.0 &  $ <    76 $ &  $ <    22 $ &  \nodata &  \nodata \\
SR 9              & 73.8 &  \nodata &  $ <    28 $ &  \nodata &  \nodata \\
SR 12             & 37.5 &  $ <    62 $ &  $ <    26 $ &  \nodata &  \nodata \\
SR 20             & 8.9 &  \nodata &  $ <    28 $ &  \nodata &  \nodata \\
V853 Oph          & 1.6/53.8 &  $    52 \pm  8 $ &  $   134 \pm 12 $ &  $   320
\pm 53 $ &  \nodata \\
VV Sco            & 240 &  \nodata &  $ <    63 $ &  \nodata &  \nodata \\
WSB 3             & 75.0 &  \nodata &  $ <    34 $ &  \nodata &  \nodata \\
WSB 11            & 62.5 &  \nodata &  $ <    26 $ &  \nodata &  \nodata \\
WSB 18            & 138 &  \nodata &  $ <    46 $ &  \nodata &  \nodata \\
WSB 19            & 188 &  \nodata &  $ <    78 $ &  \nodata &  \nodata \\
WSB 26            & 150 &  \nodata &  $ <    77 $ &  \nodata &  \nodata \\
\hline
\end{tabular}
\tablenotetext{a}{Projected separations greater than 1 AU are based on
  measured angular separations and assume distances of 125 pc to $\rho$
  Oph and 160 pc to Upper Sco.  Projected separations less than 1 AU
  are from spectroscopic orbits and are projected semimajor axes
  $a\,\sin i$.}
\tablenotetext{b}{These multiple systems have pairs with projected
  separations both less than and greater than the 1 AU sample cutoff.
  For the purposes of the statistical analysis, the pairs with $a_p <
  1$ AU are ignored.}
\end{table}

\begin{table}[ht]
\caption{\sc Surveys for binaries in Sco-Oph
\label{table:binary_surveys}}
\vspace*{5mm}
\begin{tabular}{lcll}
 Survey & Separation Range\tablenotemark{a} & Technique & Target List\\
\hline
Mathieu et al.\ (1989) & 0.01 -- $\sim$ 2 AU & Optical &
X-ray selected stars, primarily \\[-1mm]
&& spectroscopy & naked T Tauri stars\\[2mm]
Ghez et al.\ (1993)  & 14--250 AU & 2.2 \micron\ speckle  & HBC
and x-ray selected stars \\[-1mm]
&&interferometry & from Walter et al.\ (1994)\\[2mm]
Reipurth \&  & 140--1700 AU & 0.9 \micron\ imaging &
H$\alpha$ emission stars from \\[-1mm]
 Zinnecker (1993) &&&  Wilking, Schwartz, \& Blackwell\\[-1mm]
&&&  (1987) and Walter (1986)\\[2mm]
Simon et al.\ (1992, 1995) & 0.7--1400 AU & Lunar occultation, & Members of
$\rho$ Oph from IR\\[-1mm]
&& 2.2 \micron\ imaging & H$\alpha$, and x-ray surveys\\[2mm]
\end{tabular}
\tablenotetext{a}{Assuming a distance of 140 pc.}
\end{table}

\begin{table}[ht]
\caption{\sc Confidence levels from two-sample tests
\label{table:asurv_results}}
\begin{tabular}{llrr}
\\
\hline
\hline\\
Groups & Region & Number    & Confidence level \\
       &        & in groups & that groups {\em differ}\\
\hline\\
Binaries with $a_p \le 50$ AU vs.\ $a_p > 50$ AU & Tau-Aur&28/33 &97 -- 98\% \\
& Sco-Oph &10/22 &89 -- 95\% \\
& Combined &38/55 &99 -- 99.8\% \\
Binaries with $a_p \le 100$ AU vs.\ $a_p > 100$ AU & Tau-Aur&37/24 &98 -- 99\%
\\
& Sco-Oph &17/15 &91 -- 95\% \\
& Combined &54/39 &99.6 -- 99.9\% \\
Binaries with $a_p \le 50$ AU vs.\ single stars & Tau-Aur&28/55 &99.5 -- 99.7\%
\\
& Sco-Oph &10/47 &93 -- 98\% \\
& Combined &38/102 &99.9 -- 99.96\% \\
Binaries with $a_p \le 100$ AU vs.\ single stars & Tau-Aur&37/55 &99.5 --
99.6\% \\
& Sco-Oph &17/47 &93 -- 97\% \\
& Combined &54/102 &99.92 -- 99.97\% \\
Binaries with $a_p > 50$ AU vs.\ single stars & Tau-Aur&33/55 &54 -- 74\% \\
& Sco-Oph &22/47 &31 -- 51\% \\
& Combined &55/102 &51 -- 75\% \\
Binaries with $a_p > 100$ AU vs.\ single stars & Tau-Aur&24/55 &22 -- 54\% \\
& Sco-Oph &15/47 &3 -- 10\% \\
& Combined &39/102 &$<1$ -- 33\% \\
\hline\\
\end{tabular}
\end{table}

\begin{table}[ht]
\caption{\sc Disk masses from various gap clearing models
\label{table:disk_masses}}
\begin{tabular}{lrrrrr}
\\
\hline\\[-4mm]
Name & Proj.\ Sep.\ & \multicolumn{1}{c}{$\lambda$} &
\multicolumn{3}{c}{Disk Mass ($10^{-3}$ \msun)} \\
& \multicolumn{1}{c}{(AU)} & (\micron) & No clearing & Gap
$0.4a$--$1.8a$ & Gap $0.2a$--$3a$\\
\hline
\hline\\
162819$-$2423 &  2.0 &   800 & $<    0.5$ & $<    0.8$ &  $<    1.2$\\
SR 20        &  8.9 &   800 & $<    2.1$ & $<    3.9$ &  $<    7.5$\\
162814$-$2427  & 19.4 &   800 & $<    0.4$ & $<    0.4$ &  $<    0.4$\\
SR 24N       & 24.6 &  1300 & $<    1.9$ & $<    2.2$ &  $<    4.1$\\
160946$-$1851  & 33.3 &   800 & $<    2.8$ & $<    3.2$ &
\nodata\tablenotemark{a}\\
SR 12        & 37.5 &   800 & $<    0.3$ & $<    0.3$ &  $<    0.2$\\
V853 Oph     & 53.8 &   800 & $    5.2$ & $    5.3$ &
\nodata\tablenotemark{a}\\
SR 9         & 73.8 &  1300 & $    1.6$ & $    1.2$ &  $    1.1$\\
Haro 1--4    & 90.0 &   800 & $    1.5$ & $    1.1$ &  $    1.1$\\
AS 205       & 211 &   800 & $   54.3$ & $   56.6$ &
\nodata\tablenotemark{a}\\
VV Sco       & 240 &   800 & $<    7.7$ & $<    7.5$ &  $<    6.4$\\
DoAr 24E     & 257 &   800 & $    0.5$ & $    0.5$ &  $    0.4$\\
\hline\\
\end{tabular}
\tablenotetext{a}{The flux from an optically-thick disk with this
  clearing is less than\\ the measured flux or flux limit, so no mass could
  be calculated.}
\end{table}

\begin{table}[ht]
\caption{\sc JCMT fluxes for other \pms\ stars
\label{table:other_data}}
\vspace*{5mm}
\begin{tabular}{lccccc}
\hline
Name & Projected & $F_\nu\ (1100\ \micron)$ & $F_\nu\ (800\ \micron)$ &
                   $F_\nu\ (450\ \micron)$ & $F_\nu\ (350\ \micron)$ \\
     & Sep\null. (arcsec) & (mJy) & (mJy) & (mJy) & (mJy) \\
\hline
\hline
BZ Sgr            & 5 &  \nodata &  $   116 \pm 26 $ &  \nodata &  \nodata \\
CoD $-$29\deg8887  & $\le 0.8$ &  \nodata &  $ <    60 $ &  \nodata &  \nodata
\\
CoD $-$33\deg7795  & single &  \nodata &  $ <   126 $ &  \nodata &  \nodata \\
CoKu Ser/G1       & 3.5 &  \nodata &  $ <   147 $ &  \nodata &  \nodata \\
FK Ser            & 1.33 &  $ <    47 $ &  $    51 \pm  8 $ &  $ <   818 $ &
\nodata \\
HD 150193         & 1.1 &  \nodata &  $   123 \pm 18 $ &  $   455 \pm 150 $ &
\nodata \\
HD 98800          &1.5/SB1/SB2  &  \nodata &  $   108 \pm 13 $ &  $ <   365 $ &
 \nodata \\
Hen 3-600         & 1.4 &  \nodata &  $    53 \pm 15 $ &  \nodata &  \nodata \\
S CrA             & 1.4 &  $   361 \pm 26 $ &  $   765 \pm 51 $ &  $  2750 \pm
383 $ &  \nodata \\
V4046 Sgr         & SB2 &  $   451 \pm 20 $ &  $   770 \pm 39 $ &  $  2042 \pm
111 $ &  $  3154 \pm 419 $ \\
\hline
\end{tabular}
\end{table}

\clearpage

\begin{figure}[ht]
\caption{Submillimeter flux vs.\ binary separation for
  Scorpius-Ophiuchus \pms\ binaries.  Filled symbols are detections,
  open symbols are 3$\sigma$ upper limits.  Squares are 1300-\micron\
  fluxes and triangles are 800 \micron\ fluxes scaled to 1300 \micron\
  assuming optically-thick emission in the Rayleigh limit.  All fluxes
  are scaled to a distance of 140 pc.  A ``T'' next to a point
  indicates that it represents the projected separation of one pair in
  a triple system.  Note the higher fluxes among binaries with $a_p >
  $ 50--100 AU\null.
\label{fig:sco-oph}
}
\end{figure}

\begin{figure}[ht]
\caption{Submillimeter flux vs.\ binary separation for
  Taurus-Auriga \pms\ binaries.  Symbols are as in Figure 1.
\label{fig:tau-aur}
}
\end{figure}

\begin{figure}[ht]
\caption{Submillimeter flux vs.\ binary separation for
  the combined sample of Taurus-Auriga and Scorpius-Ophiuchus \pms\
  binaries.  Symbols are as in Figure 1, with all fluxes scaled to a
  distance of 140 pc.
\label{fig:combined-sample}
}
\end{figure}

\begin{figure}[ht]
\caption{The same sample as in Figure 3 but with a disk-clearing model
  superimposed.  The solid line shows the flux expected from
  binary-disk systems if they clear gaps in their disks from 0.2 to 3
  times the binary separation; the dashed line shows clearing from 0.4
  to 1.8 times the binary separation.  The model shown has $T_1 =
  150$ K, $q = 0.6$, $M_d = 0.05 $ \msun, $R_d = 100$ AU, and is
  the average over a range of inclination angles.
\label{fig:clearing-model}
}
\end{figure}

\begin{figure}[ht]
\caption{The effect of different disk parameters on gap clearing
  models for binary-disk systems.  For each plot, one model parameter
  is varied while the others are held fixed.  Unless otherwise
  specified, the model parameters are $T_1 = 150$ K, $q = 0.6$, $M_d =
  0.01 $ \msun, $R_d = 100$ AU, and $\lambda = 1300$ \micron; all
  models are averaged over a range of inclination angles. Line types
  are as in Figure 4.  (a) Varying disk temperature normalization
  $T_1$.  (b) Varying temperature exponent $q$.  (c) Varying disk mass
  $M_d$.  (d) Varying wavelength of calculated flux.  Note that the
  vertical scale is not the same on all plots.
\label{fig:4-clearing-models}
}
\end{figure}

\begin{figure}[ht]
\caption{IRAS 60 \micron\ flux vs.\ projected binary separation for
  \pms\ binaries.  Filled squares are detections, open squares are
  upper limits, and a ``T'' indicates that a system is part of a
  triple.  Most binaries are detected, indicating the presence of
  circumstellar disks in most systems.  There is no evidence for a
  relationship between 60 \micron\ flux and binary separation.
\label{fig:iras}
}
\end{figure}

\end{document}